# Quantifying the direct and indirect impact of COVID-19 vaccination: evidence from Victoria, Australia

Lixin Lin[1], Haydar Demirhan[1], Peter Eizenberg[2], James M. Trauer[3], Lewi Stone[1*]

[1] Mathematical Sciences, School of Science, RMIT University, Melbourne, Australia

[2] Doctors of Ivanhoe, Melbourne, VIC, Australia

[3] School of Public Health and Preventive Medicine, Monash University, Melbourne, Australia

**Abstract**

Vaccines not only directly protect vaccinated individuals but also contribute to protect the entire population via indirect herd-immunity benefits. However researchers have long struggled to quantify these indirect effects at the population level, hindering assessment of vaccination program effectiveness. We developed a new method to estimate these effects, thereby markedly improving measures of the number of infections, hospitalizations, and deaths averted by vaccination. Our population-based analysis of 6,440,000 residents of Victoria, Australia reveal strong indirect effects during the Delta outbreak (September–November 2021). By modelling a non-vaccination counterfactual, we conservatively estimate 316,000 infections were averted (95%BCI: 232k–406k), as well as 33,500 hospitalizations (95%BCI: 22.2k–46.2k), and 4,900 deaths (95%BCI: 2.9k–7.3k). These are 4.0, 7.5, and 8.0 times higher, respectively, than observed. Half of the averted infections and around one-quarter of hospitalizations and deaths were attributable to indirect protection. Homogeneous vaccination across LGAs could have reduced outcomes by ~25%.

## 1. Introduction

Vaccines emerged as a critically important public health intervention during the COVID-19 pandemic, with clinical trials demonstrating high efficacy within a year of the emergence of SARS-CoV-2 [1, 2]. In the post-pandemic period, there is growing interest in quantitatively evaluating the success and impact of vaccination at the population level [3, 4] including quantifying the number of infections, hospitalizations and deaths averted by vaccination. Such analyses are essential not only for understanding the effects of the COVID-19 vaccination, but also for developing frameworks to quantify and anticipate the impact of future pandemics.

Evaluating vaccination programs is complicated by the fact that population-level protection operates through both direct and indirect mechanisms [3, 5]. The direct effect, which is relatively straightforward to quantify, refers to the protection conferred to individuals by their own vaccination history [6-8]. In contrast, indirect protection occurs when chains of infection are halted upstream by immunized individuals, thereby shielding individuals (both unvaccinated and vaccinated) from exposure [6-8]. This mechanism reduces the likelihood of exposure to the pathogen and contributes to the development of so-called "herd immunity." (A more targeted application of indirect protection is the cocooning strategy, in which close contacts of vulnerable individuals—such as infants—are vaccinated to block potential transmission.) However, the concealed nature of indirect effects renders them exceptionally difficult to measure.

---

[*] Author for correspondence: Lewi Stone, email: lewi.stone@rmit.edu.au

To avoid this difficulty, several recent high-profile efforts to assess COVID-19 vaccine impact have simply ignored their presence, yielding conservative estimates of infections averted that likely underestimate the vaccine's true impact. Studies in Israel [9], U.S. [10], Japan [11] and New South Wales, Australia (NSW) [12], lacking robust methods to account for indirect effects, have thus all produced conservative estimates. Some have turned to mechanistic simulation models [3, 13, 14], which depend on numerous assumptions and parameters that are difficult to validate and require complex model fitting [15-17].

More fundamentally, even identifying indirect effects in population-level data presents a major challenge. The majority of studies estimating the indirect effects of COVID-19 vaccination were not conducted at the population level. Instead, they focus on specific settings such as households and aged-care facilities in Australia (NSW) [18], England [19], Finland [20], Scotland [21], Israel [22, 23], the Netherlands [24], and the Philippines [25]. Similar analyses have been conducted for long-term care facilities in Spain [26], day-care centers in Germany [27], California state prisons [28] (all for COVID-19), and small communities in Canada [29] (for influenza). Among the few studies that have attempted to assess indirect effects at the population level, many were constrained by insufficient data on prior population immunity, thereby providing only limited evidence of indirect vaccination effects (as in Brazil [15] and Israel [30]). Others were constrained by small populations with few recorded infections, potentially leading to inaccuracies (as in Schwaz, Austria [31]). Similar challenges apply to other infectious diseases such as influenza, where evidence of population-level indirect effects remains limited, with the classic study from Japan [32] being a notable exception. Moreover, none of the aforementioned studies have evaluated the overall population-level impact of vaccination campaigns in terms of infections and deaths averted. In contrast, the empirical approach we employ addresses both limitations: it enables the detection of indirect effects in population-level data and provides a framework to quantify the broader impact of vaccination campaigns in terms of infections, hospitalizations and deaths averted.

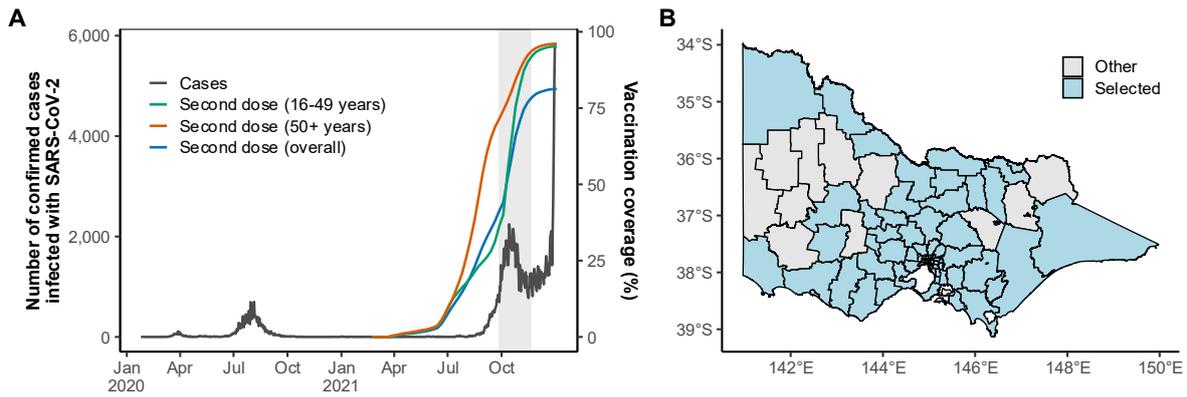

**Figure 1. Epidemic context and geographic focus in Victoria, Australia.** (**A**) COVID-19 spread in Victoria, Australia, from January 2020 to December 2021. The grey shaded band marks the study period (26 September to 21November 2021). Daily confirmed SARS-CoV-2 cases [33] (left-hand side vertical axis) are plotted alongside second dose vaccine uptake [34] (right-hand side vertical axis) in three groups: the total population (blue line), those aged 16-49 (green line), and those aged $50^+$ (orange line). The Delta wave began on 12 July 2021 and by November 2021 had grown into the largest wave Victoria would see in 2020-2021 [35]. In mid-December 2021, the Omicron variant emerged and triggered a new wave of infections [36]. (**B**) Map of Victoria showing selected LGAs included in the analysis (blue) and others (grey).

There are also strong policy reasons to estimate indirect vaccine effects. Linking vaccine coverage targets to the easing of public health restrictions is more justifiable if indirect protection can be demonstrated [37-39], whereby benefits to an individual may depend on the vaccination coverage of their surrounding community [40, 41]. This is especially relevant for COVID-19, where many vaccinated individuals—particularly the young and healthy—stood to gain relatively little personal benefit [42, 43]. For these reasons, distinguishing and quantifying direct and indirect effects remains essential for reflecting on pandemic policy decisions.

High quality surveillance data of 6,440,000 residents in Victoria, Australia (VIC) provides a unique opportunity to assess these effects during the COVID-19 period. The aggressive early control strategy in VIC resulted in very low levels of pre-existing infection-derived immunity by mid-2021 (Fig. 1A), creating an ideal environment to explore and model direct and indirect vaccine effects across Victoria's 67 local government areas (LGAs) (Fig. 1B).

## 2. Evidence of indirect protective effects

### 2.1. Study design

In this study, we focused on the Delta period from 26 September 2021 to 21 November 2021. Earlier periods were excluded due to the unavailability of vaccination coverage data at the local government area (LGA) level. Nonetheless, the selected timeframe is representative of the Delta wave, as it encompasses the majority and peak of the wave as observed in Victoria (VIC) overall (see Fig. 1A). We excluded LGAs with a population of fewer than 10k or with fewer than 10 total infections during this period, resulting in the inclusion of 67 out of 79 LGAs (see

Fig. 1B) with 6,440 thousand individuals in our analysis. Individuals were classified as FV or fully vaccinated, if they had two doses of the vaccine. Those that were not FV were classed as unvaccinated UV even if they had received a single dose, since the first dose only was insufficiently effective against the Delta variant [44]. For each selected LGA during the study period, we studied the following six subgroups: <12 years; 12-15 years; 16-49 years, FV; 16-49 years, UV; $50^+$ years, FV; and $50^+$ years, UV. Further age stratification, such as $60^+$ years, was not feasible due to limitations in the availability of age-specific vaccination coverage data at the LGA level.

The *Per capita cumulative Infection* risk, or $PI(v)$, is defined as the proportion of at-risk susceptible individuals that are infected over the study period, for vaccination level $v$. The $PI$ for each subgroup was calculated as the proportion of cumulative number of new reported infections in that subgroup (see SM1b). As only a small proportion of the total population (33k of the 6.44 million, or about 0.5%) had been infected before the study period [33], we suppose the total subpopulation is initially fully susceptible. Thus for children under 12

$$PI(v)_{<12y} = \frac{C_{<12y}}{N_{<12y}}$$

where $C_{<12y}$ is the number of reported infections in children under 12 years, and $N_{<12y}$ is the total number of children in this age group. To enhance readability subgroup subscripts are dropped henceforth. It can be seen that over the time period of a full epidemic the $PI(v)$ is closely related to what is referred to in epidemiology as the final size (or equivalently attack rate), i.e., the total proportion of the population that becomes infected [6, 45, 46].

Additionally, we determined the overall vaccination level ($v$) for each LGA, defined as the average second-dose vaccination coverage during the study period (see SM1a). Finally, we employed a Bayesian logistic model (see SM1c) to assess the empirical relationship in the observed data between $PI(v)$ and the overall vaccination level ($v$) across the selected LGAs for each subgroup.

## 2.2. Negative correlation between *PI* and vaccination levels ($v$)

**Rationale**. Clearly, if there were no indirect protective effects of vaccination, community vaccination coverage would have no effect on unvaccinated individuals. In this case, we should expect the *PI* of unvaccinated individuals to remain constant as local vaccination coverage ($v$) increases. Graphically, this relationship would be represented by a horizontal line when plotting the *PI* of unvaccinated individuals against $v$. In the absence of indirect effects, vaccinated individuals would still receive direct protection, resulting in a lower *PI* compared to the unvaccinated. However, since this direct protection for an individual is substantially determined by vaccine efficacy, the *PI* of vaccinated individuals should also remain unchanged, regardless of local vaccination coverage, and would again graphically be represented as a horizontal line. Similar considerations should apply for any sufficiently homogeneous population subgroup with respect to vaccination history. This method that allows detection of indirect effects has been used in several studies [15, 17, 30, 47].

Across all subgroups, including both fully vaccinated and unvaccinated individuals of different ages, the *PI* decreased with increasing overall vaccination level ($v$) across the 67 selected LGAs in VIC during the study period. This trend is seen clearly for each of the six subgroups in Fig. 2. The data in each subgroup was fitted with a Bayesian logistic model (see SM1c) to capture potential non-linear relationships between $PI$ and the overall vaccination level (see Fig. 3C and S3). The model estimates the main parameters: $L$ (the upper asymptote of the logistic curve) and $k$ (the logistic growth rate or steepness of the curve). Table S1 shows the estimated parameters from the fitted Bayesian logistic model for each of the six subgroups. In Fig. 2, the brown points denote the observed $PI$ for each subgroup at each LGA, while the black curve shows the mean $PI$ estimate, with a blue shaded 95% Bayesian credible interval (BCI) to indicate uncertainty from the logistic model fitting.

All six fitted logistic models had slope parameter $k<0$ (and all with statistical significance) indicating a strong negative association between $PI$ and vaccination $v$ for all subgroups. For example, the "<12 years" subgroup, had estimated parameter had $k=-12.3$ (mean, 95% BCI, -16.6 to -8.1). As shown in Fig. 2 and Table S2, increasing the vaccination level from 40% to 60% reduced $PI$ by approximately 60-90% across all subgroups, with $PI$ declining from 3.3% (mean) to 0.6% (<12 years), 3.0% to 0.6% (12-15 years), 0.7% to 0.3% (16-49 years, FV), 4.1% to 0.9% (16-49 years, UV), 0.7% to 0.2% ($50^+$ years, FV), and 3.4% to 0.5% ($50^+$ years, UV). The greatest absolute declines in $PI$ (approximately 3 percentage points) occurred in the "16-49 years, UV" and "$50^+$ years, UV" subgroups, while the "<12 years" and "12-15 years" subgroups showed reductions of approximately 2.5 percentage points.

Comparisons across subgroups further revealed that differences in $PI$ were more strongly driven by vaccination status than by age group. At $v=40\%$, $PI$ was estimated to be approximately 83% lower in fully vaccinated individuals compared with their unvaccinated counterparts in the 16-49 age group, and 80% lower in the $50^+$ age group. At $v=60\%$, the corresponding estimated reductions were approximately 71% and 66%, respectively.

Additional analysis (see SM1c, Figs. S1 and S2), using color gradients to indicate poverty rates and sex ratios, revealed no clustering among the 67 selected LGAs. As such, we found no strong evidence to suggest that demographic and socio-economic factors notably influenced the observed levels of $PI$. This is discussed in more detail in section SM1c.

For comparison purposes, section SM1d applied a linear model to fit the observed data for each subgroup. The fitted slope parameter (see Fig. S5) was statistically significant across all subgroups, supporting the existence of indirect effects of vaccination.

In summary, these results indicate a strong and robust negative association between the per capita cumulative risk of infection ($PI$) and the overall vaccination level ($v$), i.e., $PI$ decreased as $v$ increased, regardless of vaccination status or age group. This identifies the strong impact of herd immunity effects. To our knowledge, only one other study in Brazil [15] has demonstrated an indirect effect of vaccination on vaccinated populations.

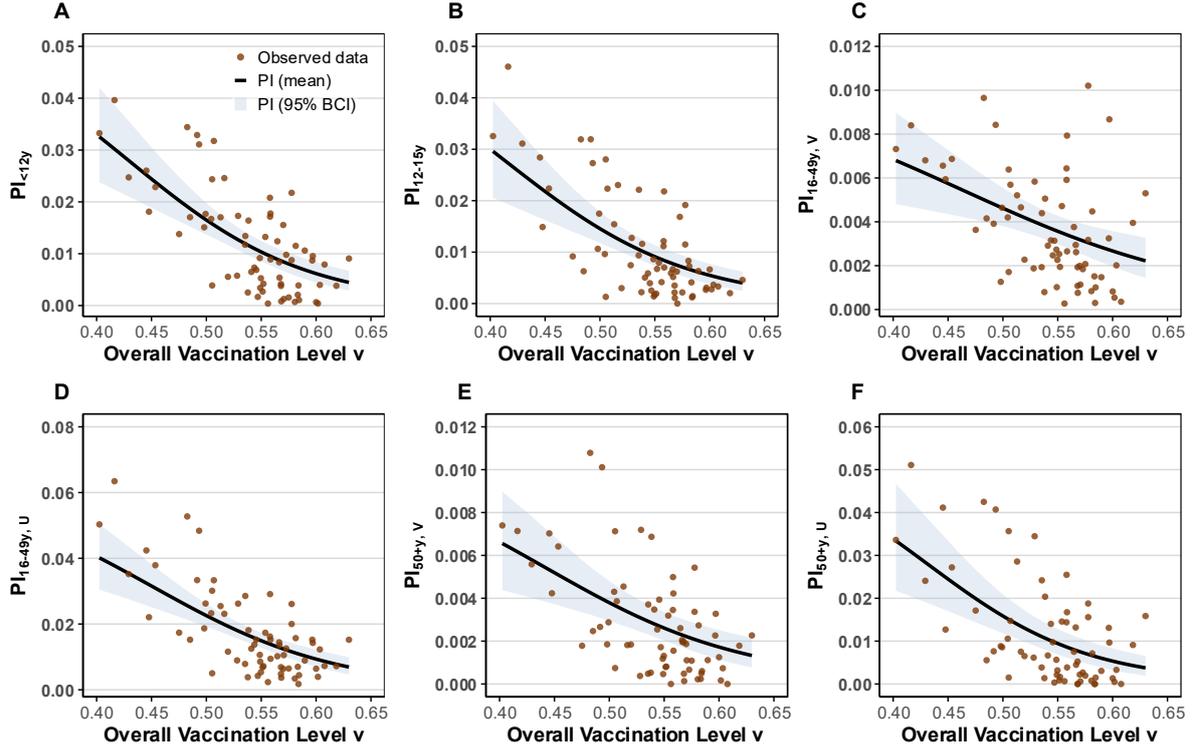

**Figure 2. Estimated per capita cumulative risk of infection ($PI$) by overall vaccination level ($v$), across six subgroups stratified by age and vaccination status during the Delta period (September 26 to November 21, 2021). Panels are: (A)** <12 years; **(B)** 12-15 years; **(C)** 16-49 years, FV (fully vaccinated); **(D)** 16-49 years, UV (unvaccinated); **(E)** $50^+$ years, FV; **(F)** $50^+$ years, UV. Each panel shows the observed data (brown dots), the mean (black curve), and the 95% BCI (blue shaded area) of the model fit to the data. A consistent decrease in $PI$ with increasing $v$ was observed across all subgroups, with the largest absolute reductions in the "16-49 years, UV" subgroup and the "$50^+$ years, UV" subgroup.

## 3. Approaches to modelling counterfactual non-vaccination scenario ($v=0$).

An important goal for this work is to estimate the number of Victorians that would have been infected had there been no vaccination at all ($v=0$). This is the hypothetical counterfactual scenario that has to be determined from observed data of the epidemic obtained when the vaccination program was in place ($v>0$) and in which hidden herd-immunity effects were operational. We need to recreate a hypothetical $v=0$ scenario in which the two-dose vaccination campaign did not occur, the first dose was ineffective or provided insufficient protection against the Delta variant [44], and in which herd-immunity effects have been untangled and removed.

The challenge in simulating counterfactual epidemic scenarios ($v=0$) using only partial data from observed conditions ($v>0$) lies in the uncertainty of the relationship between $PI(v)$ and $v$. Deriving a theorical expression for $PI(v)$ is important. Under the assumption of an imperfect vaccine, the interest will center on the infectivity of the unvaccinated, and hence we seek an expression for $PI_U(v)$ i.e., the per capita cumulative risk of infection among unvaccinated susceptible individuals for any vaccination level $v$. By learning the characteristic patterns of $PI_U(v)$ curve derived from an SIR-type model, we developed and evaluated a new method for counterfactual estimation.

A key contribution in this work is the recognition that this counterfactual (*v*=0) scenario has an epidemic with final size that is given by $PI_U(0)$, and reflects the "risk of infection" for the unvaccinated susceptible. Below, we evaluate the target methods for estimating $PI_U(0)$ by comparing their estimates with 'ground truth' values from an SIR-type model. The most reliable method is then selected to model the non-vaccination scenario (***v*=0**) in Victoria.

**3.1. SIR modelling approach**

The standard SIR model is widely used to describe the spread of infectious diseases within a population. In the standard SIR model, individuals in a population are classified into one of three states: susceptible (*S*), infected (*I*), or recovered (*R*). Here, we extend the standard SIR model by introducing two subpopulations, vaccinated and unvaccinated. Consider that at time $t = 0$, a one-time vaccination campaign begins. Initially, a fraction $I(0) = p$ ($0 < p < 1$) of the population is infected, and a fraction $v$ ($0 \leq v \leq 1$) of the remaining susceptible individuals is vaccinated with a "leaky" vaccine having efficacy $k$ ($0 < k < 1$). Similar to Lin et al [6], and Shim and Galvani [48], the differential equations representing the $S_U S_V I_U I_V R_U R_V$ model are shown below:

$$\frac{dS_U}{dt} = -\beta S_U (I_U + I_V),$$

$$\frac{dS_V}{dt} = -(1-k)\beta S_V (I_U + I_V),$$

$$\frac{dI_U}{dt} = \beta S_U (I_U + I_V) - \gamma I_U, \qquad (1)$$

$$\frac{dI_V}{dt} = (1-k)\beta S_V (I_U + I_V) - \gamma I_V,$$

$$\frac{dR_U}{dt} = \gamma I_U,$$

$$\frac{dR_V}{dt} = \gamma I_V,$$

where $S_U$, $S_V$, $I_U$, $I_V$, $R_U$, and $R_V$ denote the proportion of individuals who are susceptible (unvaccinated), susceptible (vaccinated), infectious (unvaccinated), infectious (vaccinated), recovered (unvaccinated), and recovered (vaccinated), respectively. The initial conditions are: $S_U(0) = (1-p)(1-v)$, $S_V(0) = (1-p)v$, $I_U(0) = p$, $I_V(0) = R_U(0) = R_V(0) = 0$. The basic reproductive number is $R_0 = \beta/\gamma$.

Using the above SIR-type model, it is possible to learn about the characteristics of $PI_U(v)$ analytically (see also SM1f) which assists in obtaining an estimate for the counterfactual epidemic $PI_U(0)$.

Denote $Z(v)$ as the final size of the epidemic in the entire population, i.e., the total proportion of the population that becomes infected [6, 45, 46]. Since these infections all derive from susceptible individuals, who constitute a proportion $S_U(0) + S_V(0)$ of the entire population, the per capita cumulative risk of infection among all susceptible individuals, $PI(v)$, is given by:

$$PI(v) = \frac{Z(v)}{S_U(0) + S_V(0)} = \frac{Z(v)}{1-p} \quad (2)$$

, where $Z(v) = (1-p)\left[1 - (1-v)e^{-R_0(Z(v)+p)} - ve^{-(1-k)R_0(Z(v)+p)}\right]$. (Full details in section SM1d.) Our analysis in SM1f shows that the epidemic fueled by the unvaccinated proportion of the susceptible population, $S_U(0)$, results in a final size $Z_U(v)$. This leads to the key formula:

$$PI_U(v) = \frac{Z_U(v)}{S_U(0)} = 1 - e^{-R_0(Z(v)+p)} . \quad (3)$$

Note that $PI_U(0)$ loosely reflects the risk of infection to susceptibles in an unvaccinated population.

The $PI_U(v)$ curves for the unvaccinated of the SIR-type epidemic model are plotted as a function of $v$ in Fig. 3A and B (grey lines) with the basic reproductive numbers set at $R_0$=1.5 and 4.0 respectively. The curves were generated from Eqn. (3), assuming an initial infected proportion $p$=0.01 [6, 49]. The vaccine was assumed to be 'leaky' with efficacy $k$=0.85 [44], as estimated for Pfizer, the dominant vaccine during our study period in Victoria, Australia [50]. Because the curve has a natural point of inflection (see SM1f), it declines with $v$ at low levels of vaccination but more rapidly as $v$ increases, after which it plateaus for larger $v$. These features, especially the point of inflection, motivated working with the logistic model (see section 3.2) and are also taken advantage of in formulating our new vaccination program in section 5.

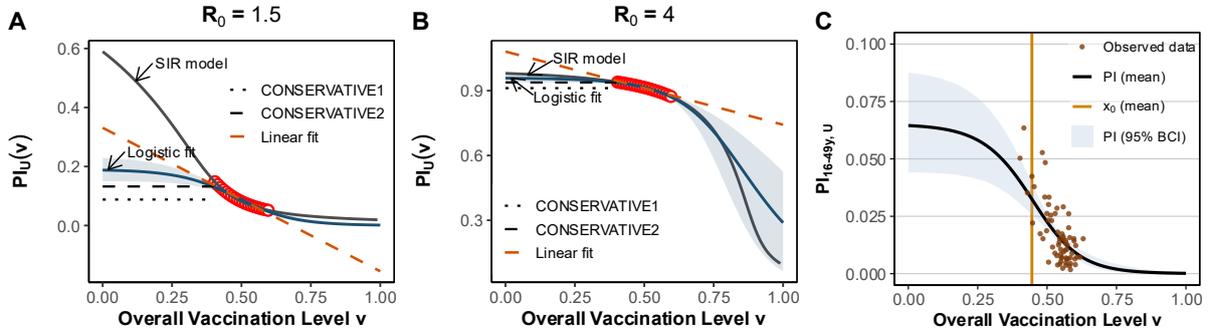

Figure 3. Estimating $PI_U(0)$ in unvaccinated individuals. A) $R_0$=1.5, and B) $R_0$=4. Panels A and B compare logistic and other approaches with SIR model. Red open circles represent the $n$=20 simulated ground truth data generated from the SIR-type model (Eqn. (3)). Different models are used to fit these points and extrapolate to find $PI_U(0)$. Grey line: $PI_U(0)$ generated by SIR model (Eqn. (3)). Blue line: posterior predictive mean of the logistic model fit. Shaded blue area: 95% BCI of logistic mean (parameter priors specified in SM1c and SM1e). Orange dashed line: linear fit. Black dashed line: CONSERVATIVE2 threshold. Black dotted line: CONSERVATIVE1 estimate for $PI_U(0)$. C) Estimated cumulative risk of infection ($PI$) as a function of vaccination level ($v$) for unvaccinated individuals aged 16-49 in 67 LGA's in Victoria during the Delta period. The observed data (brown dots), the mean (black curve), and the 95% BCI (blue shaded area) of the model fit to the data. The vertical orange line marks the mean estimated inflection point $x_0$.

To evaluate the target methods (sections 3.2 and 3.3) for estimating $PI_U(0)$, we used the $PI_U(v)$ curve from the SIR-type model (grey line, Fig. 3A, $R_0$=1.5) to generate reference model test data for 20 different hypothetical LGAs, each with a vaccination coverage levels $v$ varying between 0.4 to 0.6, similar to Victoria. We read off $PI_U(v)$ for the unvaccinated individuals for each of the LGA's, as determined by their associated $v$. These are the 20 red open circles plotted in Fig. 3A. The same data points were then used to fit four models described below, in order to estimate $PI_U(0)$. Importantly, Fig. 3A also indicates that all estimates for $PI_U(0)$ presented here are far smaller and more conservative than that given by the SIR-type model ground truth at $v$=0, namely $PI_U(0)$=59% for $R_0$=1.5.

### 3.2. Bayesian logistic modelling approach

A realistic and conservative estimate for $PI_U(0)$ can be obtained through fitting a logistic model (see SM1c and SM1e for details), as shown in Fig. 3A&B. For each test data (LGA) $i$, the observed $PI_U$ value $y_i$ was modeled as

$$y_i \sim Beta(\mu_i \varphi, (1 - \mu_i)\varphi), \tag{4}$$

with $\mu_i = \frac{L}{1+exp\{-k(x_i - x_0)\}}$, where $x_i$ is the overall vaccination level ($v$). $L$ is the upper asymptote (close to the expected $PI_U$ at zero vaccination coverage), $k$ is the slope parameter, $x_0$ is the inflection point of the curve, and $\varphi$ is the concentration parameter. Conservative priors for $L$ were set based on the maximum observable $PI_U$ value, while priors for $k$, $x_0$, and $\varphi$ are given in SM1c and SM1e.

The model was already used to fit the data in Fig. 2 when approximating the infection risk function. It is immediately obvious that for high $R_0$, as in Fig. 3B, the logistic provides an excellent fit to the SIR model's curve of $PI_U(v)$ in the regime $v$<0.75, and this is true of $PI_U(0)$ as well. In particular it captures the flat plateau seen at low levels of $v$.

Importantly, for lower levels of $R_0$=1.5, the estimated value of $PI_U(0)$ was constrained by the upper asymptote parameter $L$ of the logistic model. The logistic function has a plateau at low levels of vaccination ($v$=0) and also at very high levels ($v \simeq 1$) as shown in Fig. 3A (grey) and Fig. S6A-C (red). At intermediate levels of $v$, the logistic curve captures the general convex/concave/point-of-inflection pattern of the SIR-type model and serves as a practical approximation.

Within our Bayesian modelling framework, the conservative prior on the upper asymptote parameter $L$ deliberately restricts the $PI_U(0)$ estimate. This constraint is not due to any limitation of the logistic function itself, but is a deliberate modelling choice to avoid overestimating infection risk at zero vaccination ($v$=0), that is, $PI_U(0)$. The resulting fitted curve quickly plateaus (left of $v_{min}$=0.4 in Fig. 3C and S3) rather than linearly approximating the data that could lead to unrealistic extrapolation. The behavior reflects both the trend in the data and the intended conservatism of the estimate. Accordingly, this approach provides a more robust and credible estimate of $PI_U(0)$ (see SM1c, e).

## 3.3. Other modelling approaches

**CONSERVATIVE2:** To completely avoid extrapolation, an estimate for $PI_U(0)$ can be chosen as the maximum estimated value for $PI_U(v)$, from the logistic regression fit in Fig. 2. That is, if $v_{min}$ is the smallest observed vaccination level amongst the 67 selected LGA's, the estimate for $PI_U(0)$ is taken to be the logistic estimate of $PI_U(v_{min})$. In this case, $v_{min}= 0.4$ (see Table S2). If the SIR model is a reflection of the true epidemic (see Fig. 3A, grey line), the logistic $PI_U(v_{min})$ will surely be smaller than the true $PI_U(0)$. Thus setting $PI_U(0)=PI_U(v_{min})$ will give a conservative estimate for both $PI_U(0)$ and the averted number of infections.

**Other Models:** For comparison, we also evaluated two additional models, both of which proved inferior—full details and results are discussed in the SM1e. The model of Haas et al. (2022) [9], which is widely used and referred to here as **CONSERVATIVE1**, neglects herd-immunity effects in the observed data when estimating $PI_U(0)$, thereby undermining its accuracy and biasing it conservatively [51]. A **LINEAR** model was also analyzed but relies purely on extrapolation, and found to be least reliable.

**Bayesian Logistic model:** Results presented in SM1e suggest that the LOGISTIC estimates are closest to the reference SIR-type model, while remaining conservative, and is much closer than two other conservative approaches attempted: CONSERVATIVE1 and CONSERVATIVE2. Therefore, given the credible intervals provided by the Bayesian logistic approach and the better formulated rationale, the latter will be taken as the method of choice in this paper.

## 4. Averted outcomes by vaccination in Victoria

### 4.1. Estimating infections averted

The previous results are now made use of in estimating the number of infections averted by vaccination over study period in Victoria. To do so, we first estimate from the observed data how many infections would have occurred under the counterfactual scenario with $v=0$, which is exactly $PI_U(0)$.

For the Victoria data, we fitted the Bayesian logistic model to the per capita cumulative risk of infection (*PI*) against overall vaccination level (*v*) across 67 selected LGAs for six subgroups (see SM1c). The fit for the 16-49 unvaccinated group is shown in Fig. 3C. The fits were used to estimate $PI_U(0)$ at $v=0$ for each age group, as shown in Fig. S3. This assisted in modelling the counterfactual non-vaccination scenario ($v=0$), as described in section 3. The number of individuals infected in the counterfactual $v=0$ scenario can be easily estimated by reading off $PI_U(0)$ for each panel/age-group, and scaling by the population size. For the 16-49 year age-class this would be:

$$\#\text{Infections}_{\text{Counterfactual}} = N_{16-49y} \times PI_{16-49y,U}(0)$$

Unlike previous studies [9-12] that used the CONSERVATIVE1 approach, our estimates of infection risk $PI_U(0)$ at $v=0$ explicitly account for the full impact of vaccination including indirect effects, and can therefore be considered an advance.

The number of infections averted is easily found by subtracting the observed number of infections from this counterfactual estimate:

$$\text{\#Infections}_{\text{Averted}} = \text{\#Infections}_{\text{Counterfactual}} - \text{\#Infections}_{\text{Observed}}$$

(See also section SM2a.)

Based on the LOGISTIC fits, the total number of infections under the counterfactual non-vaccination scenario ($v=0$) is estimated to be approximately 395,000 (6.1% of the population) (mean, 95% BCI, 311k to 485k) (see Table 1), compared to the actual number of infections, which was approximately 79,000 (1.2%). Thus, the vaccination campaign was estimated to have averted approximately 316,000 (4.9%) infections (mean, 95% BCI, 232k to 406k). This estimate remains conservative due to the intentionally cautious prior set on the upper asymptote (parameter $L$) of the logistic model (see SM1c). Of the infections averted, an estimated 48.2% (mean) would have occurred in those aged 16-49, 35.5% in those aged $50^+$, and 16.3% in those under 16 (Table S4). These proportions largely reflect the age structure of the VIC population [52], as the estimated $PI_U$ at $v=0$ does not differ greatly across age groups. Table S3 also compares these results to the estimates from other approaches (CONSERVATIVE1, CONSERVATIVE2, and LINEAR). Clearly, CONSERVATIVE1 offers an extremely conservative estimate. CONSERVATIVE2 produces a somewhat conservative figure as well, approximately half the size of the LOGISTIC estimate. The LINEAR approach produces a higher estimate, indicating that around 11% of the population would become infected under the non-vaccination scenario.

**Table 1.** Number of infections, hospitalizations and deaths in thousands under a counterfactual non-vaccination scenario ($v=0$). These are estimated by the LOGISTIC approach used for $PI_U(0)$. Numbers are in thousands (i.e., 79 represents 79,000).

**Counts (mean&95%BCI) in Thousands**

| Outcome | Actual | Estimated (no vaccination) | Total averted | Direct averted | Indirect averted |
|---|---|---|---|---|---|
| **INFECTIONS** | 79 | 395 (311 to 485) | 316 (232 to 406) | 170 (116 to 227) | 146 (107 to 188) |
| **HOSPITALIZATIONS** | 4.5 | 38.0 (26.7 to 50.7) | 33.5 (22.2 to 46.2) | 24.8 (16.3 to 34.5) | 8.7 (5.7 to 12.0) |
| **DEATHS** | 0.6 | 5.6 (3.5 to 7.9) | 4.9 (2.9 to 7.3) | 3.6 (2.0 to 5.4) | 1.3 (0.8 to 1.9) |

At very low overall vaccination level ($v \approx 0$) indirect effects of vaccination are minimal and the difference in $PI$ between fully vaccinated ($PI_V$) and unvaccinated individuals ($PI_U$) primarily reflects the direct protective effect of vaccination against infection (see Lin et al. [6]). Accordingly, the direct vaccine efficacy of the two-dose regimen is defined as the relative reduction in $PI$ for fully vaccinated individuals compared with unvaccinated individuals, calculated as the proportional difference in $PI$ at $v=0$ (see Eqn. (S12) in section SM2a). Based on the LOGISTIC model estimates of $PI_U, PI_V$ at $v=0$ (see Table S2) and using Eqn. (S12), direct vaccine efficacy against infection was estimated to be 82.5% (mean, 95% BCI, 71.6% to 90.4%) for individuals aged 16-49 and 79.9% (mean, 95% BCI, 65.0% to 89.6%) for individuals aged $50^+$.

The number of infections averted through the direct effect of vaccination was estimated by multiplying the reduction in infection risk attributable to direct vaccine efficacy (i.e., $PI_U(0) - PI_V(0)$) by the number of fully vaccinated individuals (see SM2a). As shown in Table 1, according to the LOGISTIC fit, of the estimated 316,000 (4.9% of the population) (mean, 95% BCI, 232k to 406k) infections averted by vaccination, approximately 170,000 (2.6%) (mean, 95% BCI, 116k to 227k) were due to direct effects and the remaining 146,000 (2.3%) (mean, 95% BCI, 107k to 188k) were due to indirect effects.

**4.2. Estimating hospitalizations and deaths averted**

Estimates of the hospitalizations and deaths in the non-vaccination scenario ($v$=0) were also determined after calculating the observed subgroup-specific infection hospitalization rate (IHR), and infection fatality rate (IFR). These rates were estimated under actual vaccination conditions, aggregated across all 67 selected LGAs (Table S6). The rates were then applied to the estimated infection counts derived from the Bayesian logistic model for each subgroup giving hospitalizations and deaths under the counterfactual non-vaccination scenario, as outlined in the SM2b section. A simple calculation then gives the hospitalizations and deaths averted.

In the counterfactual non-vaccination scenario (see Table 1), our analysis estimates that 38,000 hospitalizations (mean, 95% BCI, 26.7k to 50.7k) would have occurred, compared to 4,500 in the actual scenario. Of the 33,500 averted hospitalizations (mean, 95% BCI, 22.2k to 46.2k), 77.1% (mean) were expected to occur in individuals aged $50^+$ years, while 21.9% were expected in those aged 16-49 (see Table S8). Furthermore, an estimated 5,600 deaths (mean, 95% BCI, 3.5k to 7.9k) would have occurred in the counterfactual non-vaccination scenario, compared to 600 deaths observed. Of the 4,900 averted deaths (mean, 95% BCI, 2.9k to 7.3k), approximately 97.6% would have been in the $50^+$ age group, with 2.2% in the 16-49 age group (see Table S10).

In addition, we estimated the number of hospitalizations averted through the direct effects of vaccination, comprising two components. First, vaccination reduces the severity of breakthrough infections among fully vaccinated individuals, attributable to its efficacy against severe disease, even when infection is not prevented. Second, it prevents hospitalizations that would have resulted from infections directly averted by the vaccine's efficacy against infection (see SM2b, Eqns. (S18-19)). Based on the LOGISTIC fit, see Table 1, approximately three-quarters of the averted hospitalizations were attributable to direct protection, with the remaining one-quarter resulting from indirect, transmission-blocking effects. A similar pattern was observed for averted deaths (see Table 1). For comparison, Tables S7 and S9 also present estimates derived from other methods. The CONSERVATIVE1 approach yields the most cautious (lowest) figures; CONSERVATIVE2 produces values roughly half of the LOGISTIC estimate, and the LINEAR method gives figures nearly twice as large as the LOGISTIC estimate.

**5. Avertable outcomes by vaccination in Victoria**

The vaccination campaign clearly prevented many infections through both direct and indirect effects (Table 1), but we have found that, further reductions could have been achieved by manipulating the spatial vaccination coverages to enhance extra indirect effects. As a demonstration, we modelled a uniform vaccination strategy that equalizes vaccination levels among the 67 selected LGAs without changing the total number of doses or the doses administered to each age group.

More specifically, vaccine allocation was adjusted only within the 16-49 age group to achieve a uniform overall vaccination coverage of 52.15% (the observed average for all selected LGAs), which reflects the overall level achieved across these areas. That is, LGAs with vaccination levels above 52.15% had their doses for this age group reduced, while those below received additional doses. We chose to relocate the vaccines in the 16-49 age group rather than the $50^+$ age group because the observed IHR and IFR were significantly lower in the younger group (see Table S6). This meant the relocation strategy would have a minimal individual-level impact and would avoid altering the prioritization of higher-risk age groups [53].

The estimated number of infections in each LGA was then recalculated by multiplying the estimated subgroup-specific per capita cumulative risk of infection (*PI*) at *v*=52.15%, derived from the LOGISTIC fit (Table S2), by the corresponding adjusted population size. Under this equalized strategy (see Table 2), the total estimated infections across all selected LGAs would be 61,000 (0.9% of the population) (mean, 95% BCI, 56k to 66k), which reflects a reduction of 23.2% (mean) compared to the actual 79,000 infections. Among the avertable infections, 56.8% (mean) would occur in those aged 16-49, 18.2% in those aged $50^+$, and 25.0% in those under 16 (Table S5). Additionally, this equalized strategy could avert 1,100 hospitalizations (mean, 95% BCI, 0.6k to 1.6k), representing 25.2% (mean) of the observed hospitalizations, and 140 deaths (mean, 95%BCI, 0.04k to 0.22k), representing 23.1% (mean) of the observed deaths (see Tables 2, S8-9).

**Table 2.** Estimated number of infections, hospitalizations, and deaths in thousands under a counterfactual scenario that with an equalized vaccination strategy. Numbers are in thousands.

| | Counts (mean&95%BCI) in Thousands | |
|---|---|---|
| Outcome | Actual (Unequal Strategy) | Estimated (Equal Strategy) |
| **INFECTIONS** | 79 | 61 (56 to 66) |
| **HOSPITALIZATIONS** | 4.5 | 3.4 (2.9 to 3.9) |
| **DEATHS** | 0.61 | 0.47 (0.39 to 0.57) |

It is important to realize that the 18,000 extra averted infections (mean, 95% BCI, 13k to 23k) was achieved without changing the number of vaccine doses. Thus it was achieved solely through changes in indirect vaccination effects. The phenomenon occurs because the marginal benefits of increasing vaccination coverage are greater in LGAs with low vaccination levels than in those with higher levels. To see this, consider Fig. 4 below where we suppose Victoria has two LGA's only with vaccination levels $v_1$ in LGA1 and $v_2$ in LGA2, where $v_1 > v_2$. When equal vaccination is applied their coverages are modified so that both LGA's have the same vaccination level $v_m$, while keeping the total number of doses unchanged. Fig. 4 shows there are major benefits in indirect vaccination effects for LGA$_1$ when $v_1$ is increased to $v_m$, as can be seen by the associated large drop in infection risk for all susceptible individuals [$PI(v_1)$-$PI(v_m)$]. On the other hand, when $v_2$ is decreased to $v_m$, there is a smaller increase in infection risk [$PI(v_m) - PI(v_2)$] for LGA$_2$. As a result of this manipulation in the doses, the number of overall averted infections increases due to the stronger impact of LGA$_1$. The theoretical details of a full multi-regional analysis are presented in section SM3 [54].

The above property relies on the SIR-type model, in which the per capital cumulative risk of infection for all susceptible individuals, $PI(v)$ (see Eqn. (2)), has a natural point of inflection $v^*$ defined by $PI''(v^*)=0$ (see SM1f). When the vaccination level $v$ exceeds the inflection point $v^*$, the decrease in risk of infection $PI(v)$ decelerates (Fig. 4; SM1f).

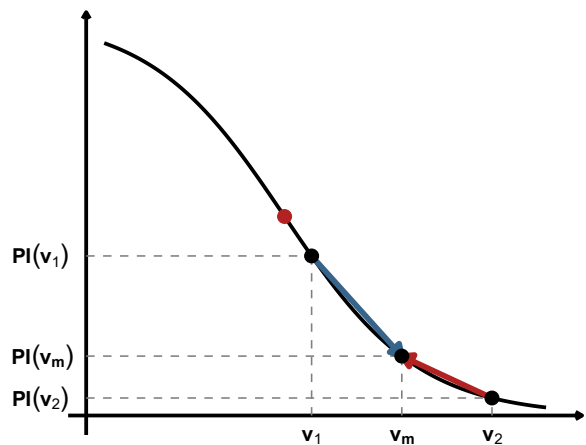

**Figure 4. Schematic illustration using a convex downward function $PI(v)$.** The black curve shows a hypothetical example of $PI(v)$, which represents the per capital cumulative risk of infection among all susceptible individuals (vaccinated and unvaccinated) to capture indirect effects of vaccine redistribution. Solid black dots mark the points with $v_1$ and $v_2$ both equidistant to $v_m$. The solid red dot marks the inflection point $v^*$ defined by $PI''(v^*)=0$. The blue and red arrows represent changes due to equal horizontal shifts of $v$ towards $v_m$ (blue: from $v_1$ to $v_m$; red: from $v_2$ to $v_m$). The associated decrease in $PI$ from $v_1$ to $v_m$ is substantially greater than the increase from $v_2$ to $v_m$, [$PI(v_1)$ - $PI(v_m)$] $\gg$ [$PI(v_m) - PI(v_2)$], illustrating the lower total risk when shifting from unequal to equal vaccination strategy.

## 6. Discussion

Observational studies from only several countries have demonstrated indirect vaccination effects at the household or population levels [15, 18-31]. However, previous population-level studies have provided relatively weak evidence of indirect effects, as in studies from Israel [30], or have had significant limitations due to prior population immunity affecting their accuracy, as in studies from Brazil [15], and also affecting the studies of Israel [30]. Victoria, Australia provides a unique context for evaluating the indirect effects of vaccination due to the almost complete absence of natural immunity in 2021 (Fig. 1A), a scenario rarely available elsewhere.

In addition, the rapid pace of vaccination and the short interval (one to two months) between vaccination and exposure to the Delta variant (Fig. 1A), recognizing that vaccine-induced immunity wanes rapidly over time [55, 56], distinguished it from other settings. Apart from these initial efforts, there are few comparable studies that examine the indirect effects of vaccination in detail at the population level, and none we know of in Australia. The study in NSW observed a reduced risk of infection in unvaccinated household contacts when the index case was vaccinated during the Delta outbreak [18]. However, the degree to which these indirect protective effects manifest was never tested beyond households to the general population.

To investigate indirect vaccination effects, we applied an innovative Bayesian logistic model inspired by SIR-derived theoretical insights to analyze how per capita cumulative risk of infection varies with overall vaccination level across 67 selected LGAs in Victoria. The results showed a statistically significant negative correlation between risk of infection and the vaccination level, providing strong empirical evidence of indirect vaccination effects. Previous observational population-level studies [15, 30] had difficulty showing this and none explored the extent to which indirect vaccination effects affected epidemic outcomes.

Our results showed that in the absence of a vaccination campaign there would have been approximately 395,000 infections (mean, 95% BCI, 311k to 485k), 38,000 hospitalizations (mean, 95% BCI, 26.7k to 50.7k), and 5,600 deaths (mean, 95% BCI, 3.5k to 7.9k), representing approximately 5.0 (mean), 8.5, and 9.0 times the actual observed values with vaccination, respectively (see Table 1). These estimates could still be considered as conservative compared to predictions derived from SIR-type modelling approach (see Fig. 3).

The analysis shows that approximately half of the infections averted (85 percent of which occurred in unvaccinated individuals) can be attributed to the indirect effects of vaccination, while about one-quarter of the hospitalizations and deaths averted can also be attributed to these indirect effects. Although the magnitude of the indirect effect on infections is substantial, it is plausible given that the effective reproductive number ($R_{eff}$) in Victoria could be low due to high vaccination level, with all LGAs above 40% and some above 60%. Using a simple SIR model, Lin et al. [6] and Eichner et al. [8] showed that when $R_{eff}$ is close to the critical threshold of one, the indirect effects of vaccination can exceed the direct effects. However, in such cases, this did not imply that the unvaccinated population was equally or better protected. In fact, unvaccinated people had a much higher observed risk of infection, hospitalization and death than those who were fully vaccinated (see Fig. 2 and Table S6). While the number of infections averted by direct and indirect effects may be similar, the number of hospitalizations or deaths averted by direct effects is significantly higher. This is because direct protection includes not only the prevention of infection, but also the reduction in severity among breakthrough infections.

Although previous studies [57, 58] have generally supported uniform vaccination strategies, they have rarely defined specific conditions under which such strategies are optimal or investigated potential exceptions. Our findings suggest that, in the context of highly infectious pathogens and when aiming for high vaccination coverage, equitable distribution is not only consistent with principles of fairness but also justified by epidemiological theory (SM3 section provides specific criteria).

In addition to regional equity, consideration should also be given to prioritizing specific age groups. For example, Bubar et al. [42] used mathematical modelling to show that prioritizing adults aged 20-49 can minimize cumulative incidence, while prioritizing those aged over 60 is more effective in reducing mortality and loss of life expectancy. Crucially, equitable allocation across regions in a multi-regional vaccination program does not necessarily conflict with age-based prioritization within each region. In optimizing Victoria's vaccination strategy, we retained age-based prioritization. Specifically, within the priority age group of 16-49 years [53], we reallocated vaccine from LGAs with higher overall vaccination level to LGAs with lower level to achieve a more balanced distribution. This reallocation strategy increased the total indirect effects, thereby reducing infections, hospitalizations, and deaths by 23.2% (mean), 25.2%, and 23.1%, respectively, across the 67 selected LGAs in Victoria (Tables S5, S11 and S12). Regional vaccination levels can be integrated as a secondary consideration, while keeping age-based prioritization the primary guiding principle.

Understanding the role and magnitude of indirect effects of vaccination is important [7, 59], but accurately quantifying these effects using observational data remains challenging. Some of the difficulties in individual case studies in Japan [60], Israel [13], and globally [3] are discussed in the SM1g section. In contrast, the empirical approach we use avoids many of these limitations and provides more directly data-driven estimates of indirect effects. The limitations of this study are also detailed in SM1g [61].

Indirect effects of vaccination are important and should not be overlooked, whether in the context of COVID-19 or other emerging infectious diseases. During the pandemic, attention has often focused on direct effects because they are easier to measure. However, indirect effects, remain an important part of how vaccines help to control outbreaks. Although concealed and harder to observe, they are just as important in understanding the full impact of vaccination programmers.

**Ethics**

Ethics approval was required to access and analyze Victoria COVID-19 infection data from the Victoria Department of Health (approval number 25603, Modelling COVID-19 and the impact of vaccination in Victoria).

**Data accessibility**

Population estimates by age and sex for each LGA of VIC in 2021 were obtained from the Australian Bureau of Statistics [52]. Data on COVID-19 vaccination coverage for all 79 LGAs of VIC were collected from the webpage provided by the Australian Government's Department of Health and Aged Care [62]. The estimated poverty rates (proportion of people living below the poverty line of 50% of median income) for each LGA in 2021 were obtained from the Victorian Council of Social Service [63]. The processed (clustered, anonymized) data on COVID-19 infections classified by vaccination status in VIC are available from the Victorian Department of Health.

**Declaration of AI use**

We have not used AI-assisted technologies in creating this article.

**Authors' contributions**

L.L.: conceptualization, formal analysis, investigation, methodology, supervision, visualization, writing—original draft, writing—review and editing; H.D.: investigation, supervision, writing—review and editing; P.E.: investigation, supervision, writing—review and editing; J.M.T.: investigation, supervision, writing—review and editing; L.S.: conceptualization, formal analysis, funding acquisition, investigation, methodology, supervision, writing—original draft, writing—review and editing. All authors gave final approval for publication and agreed to be held accountable for the work performed therein.

**Conflict of interest declaration**

We declare we have no competing interests.

**Funding**

L.S. is supported by the Australian Research Council (ARC) through grant No. DP240102585. The funder had no role in study design, data collection and analysis, decision to publish, or preparation of the manuscript.

**Acknowledgements**

None.